\documentclass[]{article}
\usepackage[]{authblk}
\usepackage[latin9]{inputenc}
\usepackage{dcolumn}
\usepackage{amsmath,amssymb,amsthm,bm,amsxtra}
\usepackage[fleqn]{mathtools}
\usepackage[x11names]{xcolor}
\usepackage[]{hyperref}%
\usepackage[normalem]{ulem} 

\newcommand{\T}{\ensuremath{\Delta t}}
\newcommand{\fobs}{\ensuremath{ B_z}}
\newcommand{\f}{\ensuremath{\{b_t}\}}
\newcommand{\unit}[1]{\ensuremath{\mathrm{~#1}}}
\newcommand{\ST}{\ensuremath{\Delta s_{\mathrm{\scriptscriptstyle T}}}}

\let\ve=\varepsilon

\begin{document}

\title{Coexistence of self-similar and anomalous scalings in turbulent small-scale solar magnetic fields.}

\author{Andrei Y. Gorobets}
\affil{Leibniz-Institut f\"{u}r Sonnenphysik (KIS), Sch\"{o}neckstr. 6, Freiburg 79104, Germany}
\author{Svetlana V. Berdyugina}
\affil{Leibniz-Institut f\"{u}r Sonnenphysik (KIS), Sch\"{o}neckstr. 6, Freiburg 79104, Germany}
\affil{Istituto ricerche solari Aldo e Cele Dacc\`o (IRSOL), Faculty of Informatics, Universit\`a della Svizzera italiana, 6605 Locarno, Switzerland}
\date{\today}%

\maketitle

\begin{abstract}
We report an evidence that self-similarity and anomalous scalings coexist in a turbulent medium, particularly in fluctuations of the magnetic field flux density in magnetized plasma of the solar photosphere. The structure function scaling exponents in the inertial range have been analyzed for fluctuations grouped according to the sign of the path-dependent stochastic entropy production. It is found that the scaling exponents for fluctuations with the positive entropy production follow the phenological linear dependence for the magnetohydrodynamic turbulence. For fluctuations with the negative entropy production, the scaling is anomalous. 
\end{abstract}

\begin{figure}[b]\center
\includegraphics[width=0.5\textwidth]{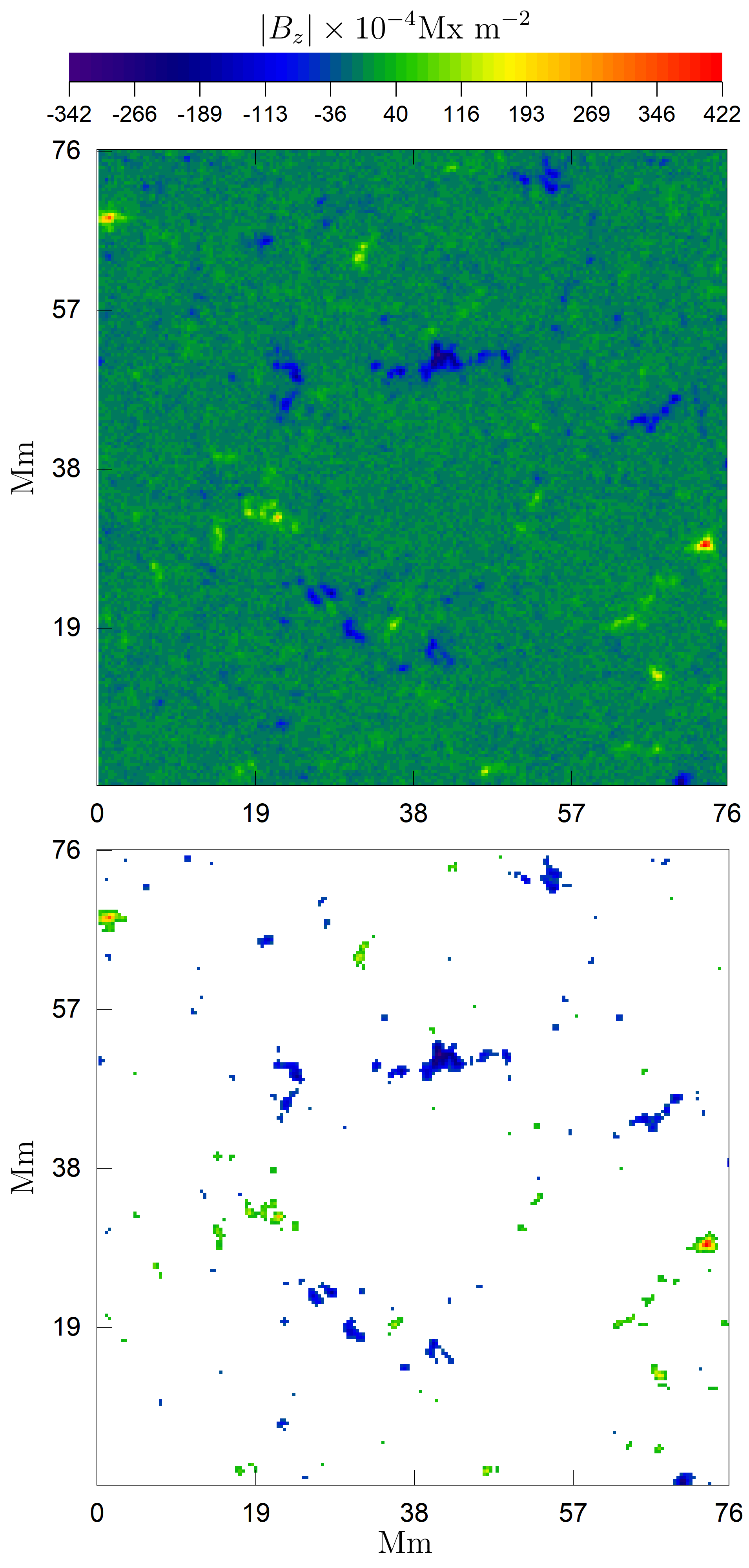}
\caption{\label{fig1} Top panel: the first magnetogram in the analyzed time-series with the FoV limited to $200^2\unit{pixels}$. The spatial sampling on the solar surface is $\approx380\unit{km}=0.5\unit{arcsec/pixel}$. The FoV {is chosen} to minimize $\fobs$ projection effects, spatial inhomogeneity of the noise, and other instrumental effects, as well as the solar differential rotation.  
Bottom panel: the same as above with the applied noise threshold cutoff of $3\sigma=3\times 10.3\times10^{-4}\unit{Mx~m^{-2}}$ \cite{liuComparisonLineofSightMagnetograms2012}. The rounded structure of MFD concentrations in the center (approx.~tens$\unit{Mm}$ scale) outlines supergranule boundaries, the so-called magnetic network \cite{rinconSunSupergranulation2018a}.
{The data analyzed here consist of $3,728,333$ stochastic trajectories from which $55\%$  are trajectories with $\ST^+$.}
}
\end{figure}

In the lower solar atmosphere (photosphere), the evolution of magnetic fields is influenced by turbulent magnetoconvective motions of plasma, especially in regions with weak fields ($\le 0.1\unit{Mx~m^{-2}}$) of the so-called "quiet Sun",
i.e. away from pores, sunspots, and their groups (active regions), where stronger magnetic fields suppress convective motions. 
The quiet Sun line-of-sight magnetic flux density (MFD) $\fobs$ is observed as a rapidly evolving, spatially intermittent (fractal) quantity in magnetic field maps (magnetograms) \cite{faurobert-schollTurbulentMagneticFields1995,consoliniScalingBehaviorVertical1999,stenfloScalingLawsMagnetic2012,janssenFractalDimensionSmallscale2003,giannattasioScalingPropertiesMagnetic2022}. 
Photospheric magnetograms (Fig.~\ref{fig1}) are recorded by space missions with a high cadence during several $11$-year solar cycles.
The range of physical parameters in the solar atmosphere provides a unique laboratory for unprecedented continuous high spatial resolution studies of dynamic magnetic phenomena \cite{schumacherColloquiumUnusualDynamics2020}. 
In this Letter, we report a first empirical evidence for a dual character of the scaling law in temporal fluctuations of $\fobs(t)$ when their statistical realizations are analysed separately according to the sign of the stochastic entropy production.

We employ an uninterrupted observation of the quiet Sun at the solar disk center obtained by the Helioseismic and Magnetic Imager (HMI) on board the Solar Dynamics Observatory (SDO) space mission \cite{scherrerHelioseismicMagneticImager2012,schouDesignGroundCalibration2012}. The analyzed time-series consists of $51,782$ magnetograms in the Fe I $617.3\unit{nm}$ line from 2019 December 11, 00:00:22~UT to 2020 January 06, 23:58:07~UT, with the instrument-fixed cadence $\T=45\,\mathrm{s}$. This is exactly  $27\unit{days}$, which is  somewhat longer than one synodic rotation period of $26.24 \unit{days}$.


The magnetogram series is considered pixel-wise as discrete, time-ordered snapshots of magnetic flux evolution in the Eulerian frame of reference. In this context, every pixel as a probe in the field of view (FoV) provides a finite-length random realization of MFD fluctuations (also called trajectory or path) 
\begin{align}\label{eq1}
\fobs(t) &:=\bigl\{\fobs(t_1),\fobs(t_1+\T), \dots,\fobs(t_1+n\T)   \bigr\}\\ \nonumber
            &=\bigl\{b_1,b_2, \dots,  b_n  \bigr\}=\f,\:\:{t\in[1,n]}, 
\end{align}\noindent
where $t$ is the local time index starting at the local origin $t_1$, $n$ is the length of the trajectory. The trajectory is a set of identically distributed, signed, non-Gaussian, random variables; sign of $b_t$ designates polarity of $\fobs(t)$ at a given time instance, and $n$ is the exponentially distributed random number. At a given pixel, the total number of trajectories $\f$ is arbitrary. It depends on: the overall observation time, a particular solar magnetic field topology within FoV, and the noise cutoff. Statistical properties of trajectories are assumed to be homogeneous in space for the quiet Sun, at least with the HMI spatio-temporal resolution
    \footnote{The empirical test of Markov property at a higher resolution in \cite{gorobetsMARKOVPROPERTIESMAGNETIC2016} revealed that granular and intergranular $\fobs$ had, to some extent, different statistical properties, which were neglected at that stage of the studies. More details of the relevant discrepancies were reported in \cite{giannattasioScalingPropertiesMagnetic2022}.}. 
Hence, trajectories of different pixels contribute to the overall statistics equally.

The nature of $\fobs$ fluctuations enables analysis of fluctuations including a measure of their irreversibility. Namely, $\T$-transitions in $\f$ obey Markov property \cite{gorobetsMARKOVPROPERTIESMAGNETIC2016}, and so allow computing trajectory-dependent (total) stochastic entropy production
\begin{align}\label{eq2}
\ST(\f)=&\ln\left[\frac{p_n\left(b_1,b_2,\cdots,b_{n}\right)}{p_n\left(b_{n},\cdots, b_{2},b_1\right)}\right] \\
=&\ln \left[\frac{p(b_1)}{p(b_{n})} \prod\limits_{k=1}^{n-1}\frac{p(b_{k+1}|b_{k})}
{p(b_{k}|b_{k+1})}\right], \label{eq3}
\end{align}\noindent
where $p$, $p_n$ and $p(b_j|b_i)$ are respectively the marginal, $n$-joint and $\T$-step conditional probability density functions (PDF). The random quantity $\ST$ is the measure of irreversibility of the trajectory, and its PDF has an exact symmetry relation, known as the detailed fluctuation theorem \footnote{For introduction and review see, for example: \cite{bustamanteNonequilibriumThermodynamicsSmall2005,
harrisFluctuationTheoremsStochastic2007,
marconiFluctuationDissipationResponse2008,
jarzynskiEqualitiesInequalitiesIrreversibility2011,
seifertStochasticThermodynamicsFluctuation2012,
klagesNonequilibriumStatisticalPhysics2013,
seifertStochasticThermodynamicsThermodynamic2019}}:
\begin{equation}
 \frac{p(\ST>0)}{p(\ST<0)} = e^{|\ST|}.\label{eq:FT}
\end{equation}

That is, the total entropy consumption, $\ST^-\equiv\ST<0$, is exactly exponentially less probable than the total entropy generation, $\ST^+\equiv\ST>0$, of the same magnitude $|\ST|$. Hereafter, the corresponding signs are placed as superscripts in notations of estimated quantities. The detailed pixel calculus and Markov property test for $\f$ at a higher spatial resolution are described in \cite{gorobetsMARKOVPROPERTIESMAGNETIC2016}. For HMI $\f$, properties of the regular Markov chains were considered in \cite{gorobetsMaximumEntropyLimit2017}, and the validity of the fluctuation theorems (including Eq.~(\ref{eq:FT})) was shown in \cite{gorobetsStochasticEntropyProduction2019}.


Henceforth, in our investigation of scale invariance of $\fobs(t)$ fluctuations due to turbulent origin, we take into account the sign of $\ST$, which defines two disjoint sets $\f^\pm$. 
The conventional method of studying manifestations of scale invariance involves an analysis of signal's self-similarity in terms of the $q$-order structure functions (SF) 
\begin{equation}\label{eq:SF}
S_{q}(\ell)\equiv\langle|\delta_\ell \fobs(t)|^q\rangle=\langle|{\fobs}({t}+\ell)- \fobs(t)|^q\rangle,
\end{equation}
\noindent
where $\delta_\ell (\cdot) $ is an increment of a turbulent quantity at two points of the flow at a distance $\ell$. The Taylor's "frozen turbulence" hypothesis connects temporal and spatial scales in measurements, so scales in Eq.(\ref{eq:SF}) are used in units of spatial distance. The solar data we investigate do not resolve all vector components of the observable/inferred quantities like photospheric velocity and magnetic fields, and consequently details of real flows are quite uncertain. However, we assume that Taylor's hypothesis is applicable for MFD of the quiet Sun  \cite{guerraSpatioTemporalScalingTurbulent2015}. For the set of 1D trajectories of a finite length, SF are computed as the ensemble average, and $\ell$ is expressed in units of the sampling interval $\T$.

The phenomelogical theory of turbulence establishes fundamental scaling relations for observable quantities, and hence defines power-law dependencies between SF. The Kolmogorov phenomenology \cite{K41} of the fully developed hydrodynamic (HD) turbulence at a high Reynolds number $R=v\ell_0/\nu$ predicts the scaling law in the inertial range $\lambda\ll\ell\ll\ell_0$:
\begin{equation}
\label{eq:K41}
\delta_\ell v \sim \ve^\frac{1}{3}\ell^\frac{1}{3},
\end{equation} 
where $v$ is the velocity, $\ve$ is the average energy dissipation rate, $\nu$ is the viscosity, and $\ell_0$ and $\lambda$ are the integral and dissipation scales, respectively.

Turbulence of a magnetized plasma is described in the framework of magnetohydrodynamics (MHD). The corresponding Iroshnikov-Kraichnan phenomenology \cite{iroshnikovTurbulenceConductingFluid1964, kraichnanInertialRangeSpectrumHydromagnetic1965}  includes the Alfv\'en wave effect of coupling between velocity and magnetic field fluctuations on small-scales by the integral-scale magnetic field $B_0$ \cite{biskampCascadeModelsMagnetohydrodynamic1994,schekochihinMHDTurbulenceBiased2022}.
At a high magnetic Reynolds number $Rm = v_Al_0/\eta$, the self-similar scaling exponents are
\begin{equation}
\label{eq:IK}
\delta_\ell v \sim \delta_\ell B \sim [\ve v_A]^\frac{1}{4}\ell^{\frac{1}{4}},
\end{equation}
where $\eta$ is the magnetic diffusivity, $v_A\equiv B_0(4\pi\rho)^{-\frac{1}{2}}$ is the Alfv\'en velocity in $B_0$, $\rho$ is the mass density, and $\ell_0 =v_A^3\ve^{-1}$.

In terms of SF, the self-similar (linear) scalings in Eqs.~(\ref{eq:K41}-\ref{eq:IK}) read
\begin{equation}\label{eq:S} S_{q}(\ell) \sim \ell^{\xi(q)}, \:\: \xi(q)=\frac{q}{m},\end{equation} with $m=3$ for HD and $m=4$ for MHD turbulence.

To cope with experimental limitations and irregularities of flows which hinder the analysis of scaling in $S_{q}(\ell)$, the concept of the Extended Self-Similarity (ESS) was proposed in Refs.~\cite{benziExtendedSelfsimilarityTurbulent1993,benziExtendedSelfSimilarityDissipation1993}. In essence, ESS is a set of the functional dependencies of SF of any order on SF of the order for which $\xi(q)=1$. Hence, for the case of MHD turbulence we focus on ESS with the relative exponents $\xi_4$ 
\begin{equation}\label{eq:ESS}
S_{q}(\ell) \sim \left[S_{4}(\ell)\right]^{\xi_4(q)}, \:\: \xi_4(q)=\frac{\xi(q)}{\xi(4)}.
\end{equation}

The linear scalings in Eq.~(\ref{eq:S}) are violated by spatial inhomogeneities of the dissipation on small scales, as said by intermittency. Thus, the scaling exponents (anomalously) deviate from the exact linear relations, as has become evident from extensive experimental and numerical studies \cite{frischTurbulence1995}. Models for intermittency differ by assumptions about statistical properties of the energy dissipation rate $\ve$, such as log-normal \cite{kolmogorovRefinementPreviousHypotheses1962}, multifractal \cite{meneveauSimpleMultifractalCascade1987}, and  log-Poisson \cite{sheUniversalScalingLaws1994,sheHierarchicalStructuresScalings1997}. The latter was revealed for the solar wind MHD turbulence \cite{grauerScalingHighorderStructure1994,politanoModelIntermittencyMagnetohydrodynamic1995} and applied for photospheric flows \cite{consoliniCharacterizationSolarPhotospheric1999}.
The "standard model" of Ref.\cite{politanoModelIntermittencyMagnetohydrodynamic1995} as the non-parametric version of the log-Poisson model for MHD turbulence
 \begin{equation}\label{eqLP}   \xi_4(q)=q/8+1-(1/2)^{q/4} \end{equation}
is used as a reference for anomalous scaling in the results presented below.

\begin{figure}[b]\center
\includegraphics[width=0.6\textwidth]{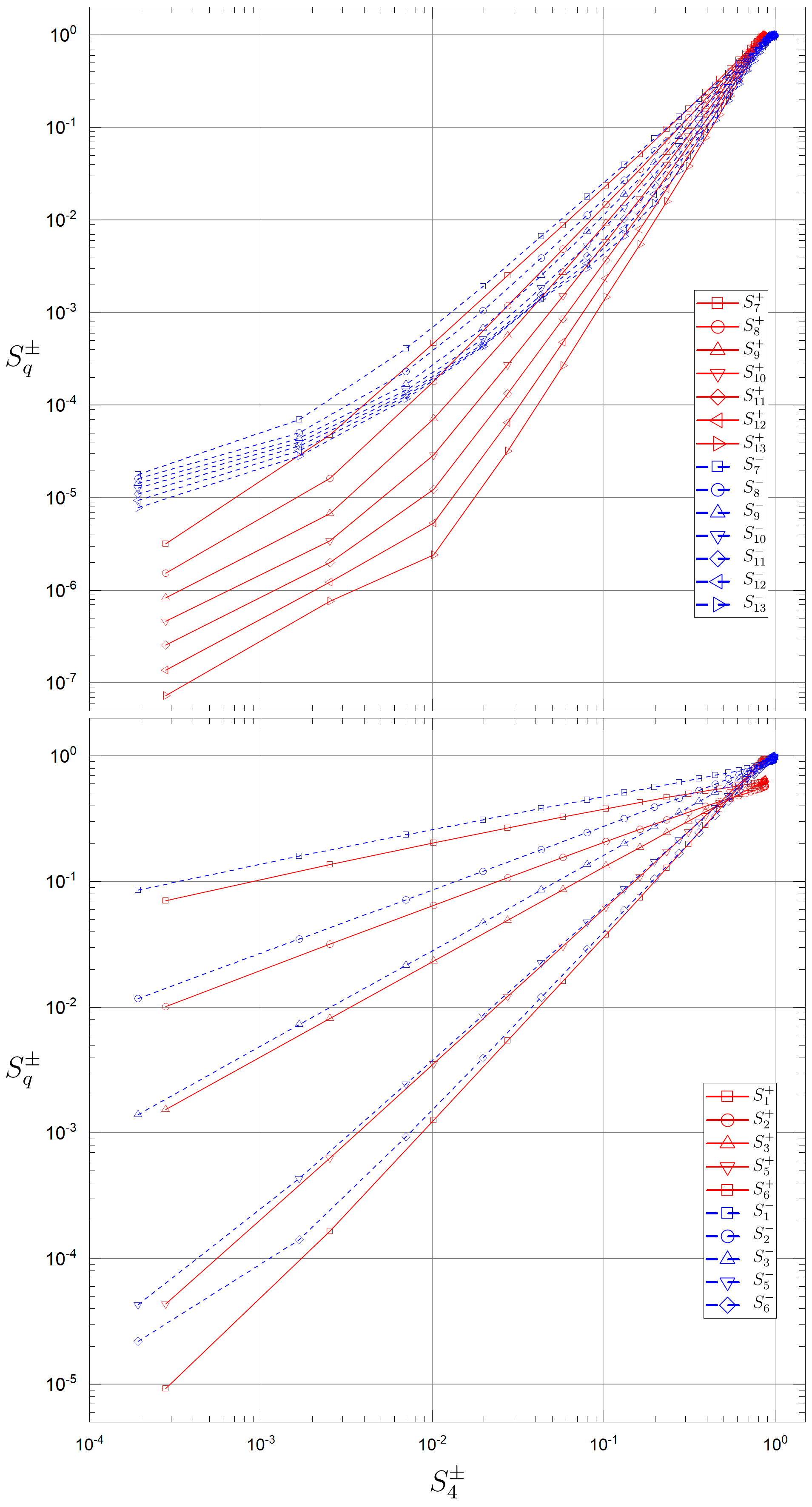}
\caption{Normalized structure functions $S_q$ of the order up to $q=13$ are shown as functions of $S_4$. {Dashed lines:} structure functions for $\ST^+$-trajectories. {Solid lines:} structure functions for $\ST^-$-trajectories. For clarity, functions are bounded by the upper limit of the inertial range.  \label{fig2}}
\end{figure}

In Fig.~\ref{fig2}, the SF scalings are shown according to Eq.~(\ref{eq:ESS}) being computed separately for two sets $\f^\pm$. The discrepancy in slopes with respect to sign of $\ST$ is clearly seen, especially for higher orders.
Following ideas from Ref.~\cite{stolovitzkyKolmogorovRefinedSimilarity1992}, the inertial range is defined as the range in which Kolmogorov's $\frac{4}{5}$ law $S_3(\ell)=-\frac{4}{5}\ve\ell$ holds. For our data, we found the inertial range to be from $15\T$ to $19\T$.

The range boundaries were modified by $\pm\T$, to compensate for a rather coarse sampling rate $\T$, because linear fits showed substantial variations with range boundaries. This modification also helps to improve statistics of fits. Therefore, an SF scaling (Eq.~\ref{eq:ESS}) in the inertial range is estimated by the set of independent linear fits within the extended inertial range $[15\T\pm\T,19\T\pm\T]$. The ultimate value of the scaling exponent $\xi_4$ is then computed as the weighted mean of $9$ exponents for every combination of the inertial range boundary variations given by $(0,\pm 1)\T$.

\begin{figure}[b]\center
\includegraphics[width=1\textwidth]{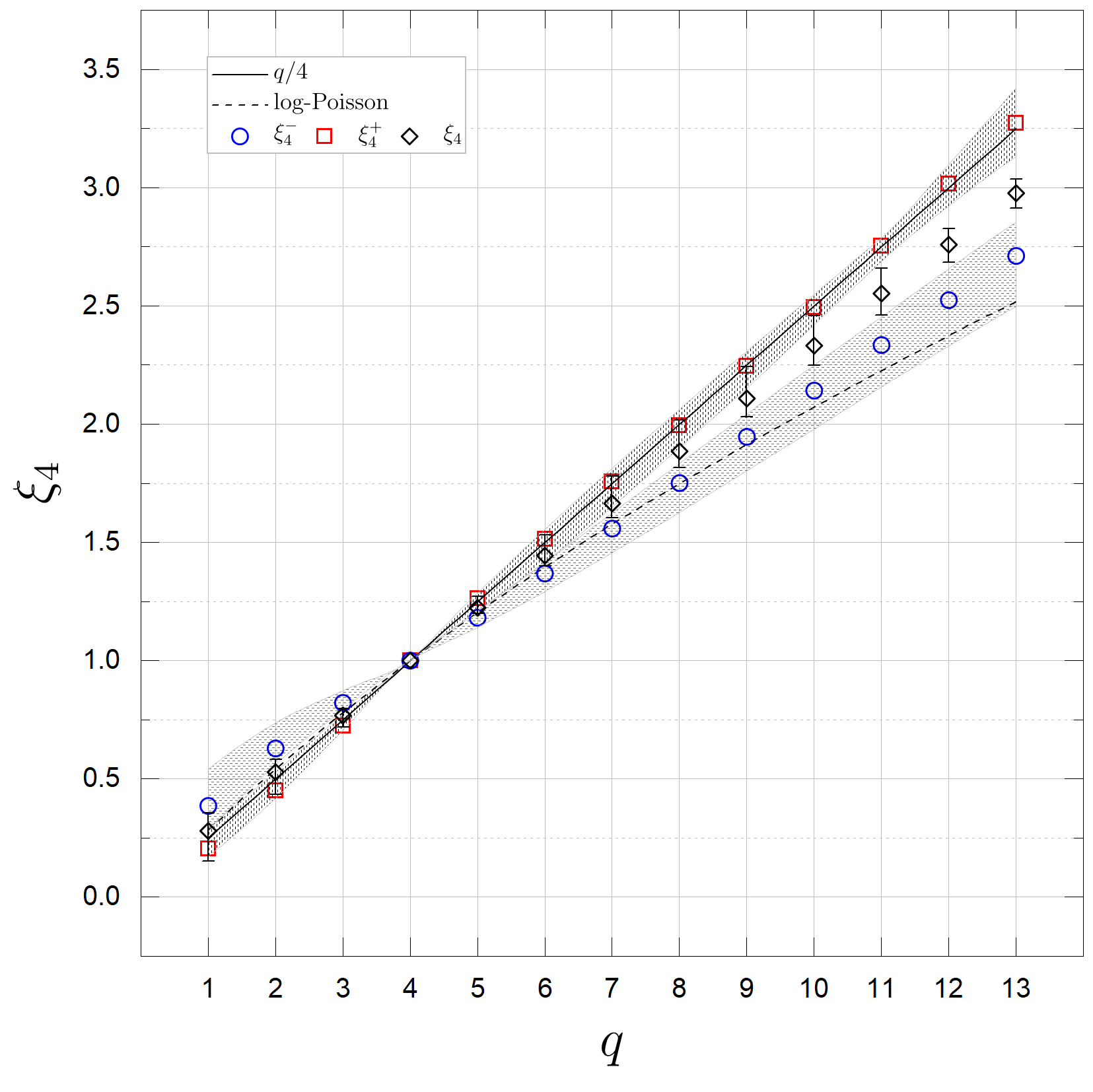}
\caption{ The relative scaling exponents $\xi_4$ as functions of SF order $q$ for sets of $\ST^+$-trajectories (squares), $\ST^-$-trajectories (circles), and the joint data set (diamonds). Bars and shadowed regions are the $99,99\%$ confidence intervals computed for $\chi^2$ merit functional. Lines are the model values with the MHD linear scaling (solid) and the anomalous scaling according to Eq.~(\ref{eqLP}) (dashed).}\label{fig3}
\end{figure}

This procedure was applied to three groups of fluctuations: $\f^\pm$ and their joint data set. The result is shown in Fig.~\ref{fig3}. Statistical robustness of the result is highlighted by the $99,99\%$ confidence level computed by the $\chi^2$ minimization. Errors of the means are smaller than symbols and not shown.

Summarizing, an anomalous scaling is the intrinsic property of the MFD fluctuations in the quiet Sun (diamonds in Fig.~\ref{fig3}). The main results is the statistically significant difference between $\xi^+(q)$ and $\xi^-(q)$. The former exhibits scaling exponents rather distinctly following the linear dependence ${\frac{q}{4}}$, in accordance with the Iroshnikov-Kraichnan phenomenology. Contrastly, fluctuations along $\ST^-$-trajectories have anomalous scaling exponents, and the curve of $\xi^-(q)$ resembles the MHD log-Poisson model (Eq.~\ref{eqLP}).
However, we note that models describing curves of $\xi(q)^-$ and $\xi(q)$ are out of the scope of the present Letter.

Following the arguments of She and Leveque \cite{sheUniversalScalingLaws1994}, 
one can interpret our finding that entropy consuming fluctuations could be related to entropy (energy) sinks which support building up of coherent structures at larger scales due to correlations induced by intermittency. Correspondingly, entropy generating fluctuations are related to dissipation processes according to the phenomenological cascade model.

To conclude, splitting measurements according to the sign of the entropy production allows detecting an unexpected coexistence of self-similar and anomalous scalings in the inertial range of turbulent small-scale photospheric magnetic fields on the Sun. 
Future numerical and experimental/observational applications of the method proposed in this Letter may advance understanding of the self-similarity in turbulent phenomena.

$$ ~~ $$
We thank Petri K\"{a}apyl\"{a} for stimulating discussions.
Solar Dynamics Observatory (SDO) is a mission for NASA's Living With a Star (LWS) program.
The Helioseismic and Magnetic Imager (HMI) data were provided by the Joint Science Operation Center (JSOC).

\providecommand{\noopsort}[1]{}

\end{document}